\documentclass[pra,twocolumn,superscriptaddress,longbibliography]{revtex4-2}
\usepackage{blindtext}
\usepackage{amsfonts}
\usepackage{amsmath,amssymb}
\usepackage{graphicx}
\usepackage{bm}
\usepackage{epstopdf}
\usepackage{url}
\usepackage{hyperref}
\usepackage{xcolor}
\hypersetup{
     colorlinks = true,
     linkcolor = blue,
     anchorcolor = blue,
     citecolor = blue,
     filecolor = blue,
     urlcolor = blue
}

\begin{document}

\title{Universal correlations along the BEC-BCS crossover}
\author{J.C. Obeso-Jureidini}
\affiliation{Instituto de F\'{\i}sica, Universidad
Nacional Aut\'onoma de M\'exico, Apartado Postal 20-364, M\'exico D.F. 01000, Mexico.} 
\author{G. A. Dom\'{\i}nguez-Castro}
\affiliation{Instituto de F\'{\i}sica, Universidad
Nacional Aut\'onoma de M\'exico, Apartado Postal 20-364, M\'exico D.F. 01000, Mexico.} 
\affiliation{Institut f\"ur Theoretische Physik, Leibniz Universit\"at Hannover, Germany}
\author{E. Neri}
\affiliation{Instituto de F\'{\i}sica, Universidad
Nacional Aut\'onoma de M\'exico, Apartado Postal 20-364, M\'exico D.F. 01000, Mexico.} 
\author{R. Paredes}
\affiliation{Instituto de F\'{\i}sica, Universidad
Nacional Aut\'onoma de M\'exico, Apartado Postal 20-364, M\'exico D.F. 01000, Mexico.} 
\author{V. Romero-Roch\'{\i}n}
\affiliation{Instituto de F\'{\i}sica, Universidad
Nacional Aut\'onoma de M\'exico, Apartado Postal 20-364, M\'exico D.F. 01000, Mexico.} 
\email{romero@fisica.unam.mx}

\begin{abstract}
We show that the long-distance behavior of the two-body density correlation functions and the Cooper-pair probability density of a balanced mixture of a two-component Fermi gas at $T = 0$, is universal along the BEC-BCS crossover. Our result is demonstrated by numerically solving the mean-field BCS model for different finite short-range atomic interaction potentials. We find an analytic expression for the correlation length in terms of the chemical potential and the energy gap at zero momentum. 
\end{abstract}
\maketitle

Equilibrium density correlation functions are fundamental for the understanding of the spatial structure of matter \cite{Landau_statistical}. They yield the next level of information beyond thermodynamics, 
unraveling the underlying arrangement in matter found at all scales, from astronomical galaxies to atomic conglomerates such as solids, liquids, or superfluid phases \cite{Stephanovich,Kusalik,Jing,Sandeep,Gadway,Shimada,Shin}. In addition, correlation functions are of fundamental relevance to characterize the order of a phase transition as they directly track density-density fluctuations \cite{Fisher,Nienhui,Fisher2,Gavai,Bernd,Gupta,Lipa,Senthil}. Our interest here is on the density correlations of the ubiquitous crossover of fermionic superfluids that goes from a Bardeen-Cooper-Schrieffer (BCS) state to a molecular Bose-Einstein condensate (BEC), as the $s$-wave scattering length is varied through a Feshbach resonance \cite{Engelbrecht,Ohashi_PRL02,Giorgini-Stringari,Ketterle-review,Feshbach-review,Chiofalo,strinatibcs}. Disregarding its charged or neutral nature, superconductor, superfluid, or molecular BEC states are the principal realizations that demonstrate how the instability associated with weak interactions develops into a many-body ground state that, besides exhibiting a second order phase transition, shows  distinctive behaviors depending on the effective attractive or repulsive nature of the interactions. 
This prominent conclusion arising from the seminal works by Leggett \cite{diat_Leggett} and Eagles \cite{Eagles_PR69}, has become a referent in the whole field of fermionic matter in its degenerate regime.
The mean-field model of BEC and BCS states represents the starting point for the understanding of more complex phenomena, as for instance, superfluids or superconductors in lower dimensions or confined in non-homogeneous environments, FFLO phases \cite{strinatibcs,Kinnunen_2018}, the cooling of neutron stars \cite{strinatibcs}, the formation of deuteron states in nuclear matter \cite{strinatibcs}, and of course, the rich diversity of systems belonging to the physics of quasiparticles and excitons \cite{Rodin-collective}.

\onecolumngrid

\begin{figure*}[h!]
\centering
\includegraphics[width=0.9\linewidth]{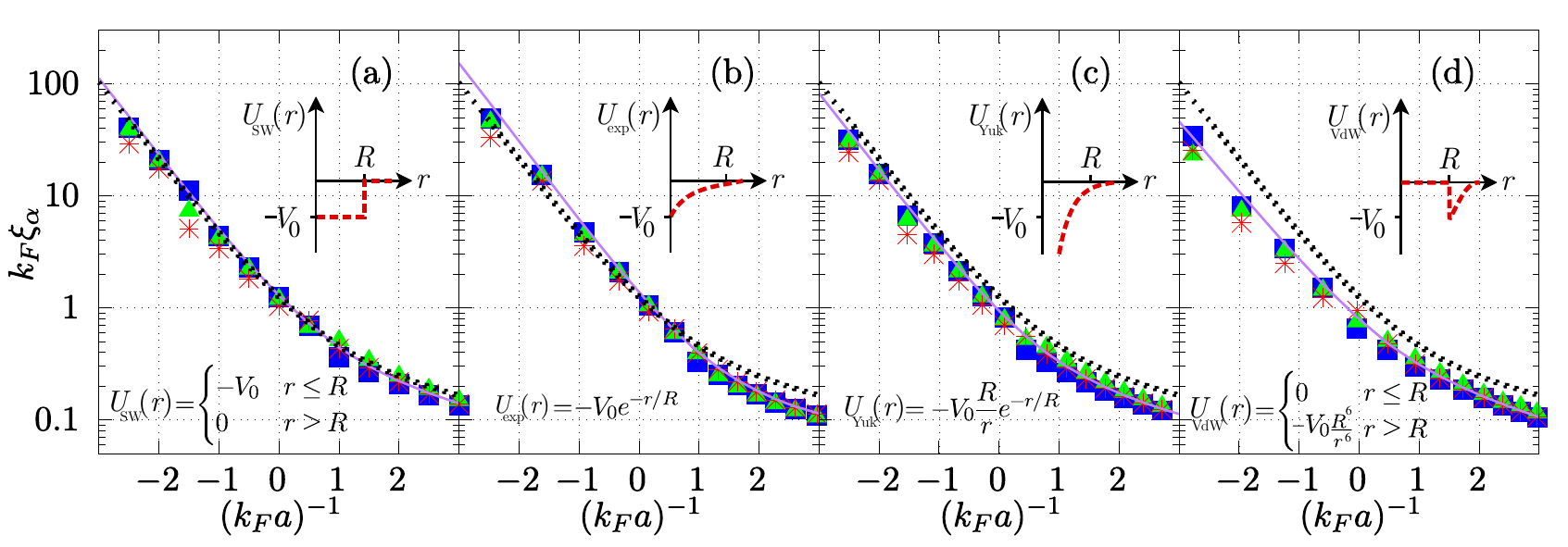}
\caption{(Color online) Large-distance correlation lengths $\xi_\alpha$ for (a) square well, (b) exponential, (c) Yukawa, and (d) Van der Waals interatomic potentials, as functions of the scattering length $a$, with $k_F$ the Fermi momentum. In each panel, squares (blue) correspond to the antiparallel spins correlation function $G_{\uparrow \downarrow}(\mathbf{r})$, triangles (green) to the parallel spins correlation function $G_{\uparrow \uparrow}(\mathbf{r})$, and asterisks (red) to the pair probability distribution function $|\phi_{\text{BCS}}(r)|^2$; see text. The solid line (purple) is the length associated with the correlation length $\xi = (\hbar^2/4m \epsilon_{\text{spec}})^{1/2}$ with $\epsilon_{\text{spec}}$ given in Eq. \eqref{espec}. In each panel the dotted (black) line corresponds to the contact interaction potential \cite{JCO1,Strinati-2022}.}
\label{fig_large-distance_correlation_length}
\end{figure*}

\twocolumngrid

An additional transcendental feature of the BEC-BCS crossover model, which is the subject of this Letter, is the universal character of the long-distance behavior of the spatial density correlations exhibited for finite short-range interatomic potentials. 
 As we will show, the correlation functions and the pair probability density can be generally expressed as an exponentially decaying function, along with an algebraic decay and an oscillatory function, regardless of the detailed features of the interaction potential at short lengths. Fig. \ref{fig_large-distance_correlation_length} shows the main finding of this Letter, namely, that the correlation length of the mean-field up-up (down-down),  up-down density correlation, and pair-probability distribution functions can be universally expressed in terms of the chemical potential $\mu$ and the energy gap at zero momentum $\Delta_0$, as $\xi = (\hbar^2/4m \epsilon_{\text{spec}})^{1/2}$, with 
\begin{equation}
\epsilon_{\text{spec}} = \sqrt{\mu^2 + |\Delta_0|^2}-\mu \>.\label{espec}
\end{equation}
This spectroscopic energy, introduced by Ketterle et al. \cite{schunckdetermination,Ketterle-review}, is the threshold energy to break a pair. It is certainly a measurable quantity and can yield physical insights for further improvements of our understanding of the BEC-BCS crossover. It is relevant to mention that characteristic lengths of the correlation functions, as well as the Cooper pair-size have certainly been the subject of several studies, see Refs. \cite{Pistolesi,Marini,Bertsch,Palestini,Chiofalo}. Our contribution to this matter is the identification of the universal lengths $\xi$ associated with the large-distance behavior while remarking its relation with an effective binding energy that is a consequence of many-body effects and independent of the interaction details.\\

The determination of the universal properties of the long-distance behavior of two-body properties of an ultracold balanced gas mixture of Fermi atoms, interacting between pairs by a  short {\it finite range} interatomic potential, see insets of Fig. \ref{fig_large-distance_correlation_length}, emerges from the Hamiltonian
\begin{equation}
 H=\sum_{\mathbf{k}, \sigma}\epsilon_{\mathbf{k}} \hat{a}_{\mathbf{k},\sigma}^{\dagger}\hat{a}_{\mathbf{k},\sigma} +
\frac{1}{2}\sum_{\mathbf{K};\,\sigma,\sigma'} U_\mathbf{k k'}\hat{a}_{\mathbf{k+ q},\sigma}^{\dagger}\hat{a}_{\mathbf{k'-q},\sigma'}^{\dagger}\hat{a}_{\mathbf{k'},\sigma'}\hat{a}_{\mathbf{k},\sigma}.\label{eq:hamiltonian}
\end{equation}
Here, $\hat{a}_{\mathbf{k},\sigma}^{\dagger}$ and $\hat{a}_{\mathbf{k},\sigma}$ are creation and annihilation Fermi operators of momentum $\mathbf{k}$ and spin component $\sigma$, identifying two different hyperfine spins states and labeled as $\uparrow$ and $\downarrow$; the kinetic energy is $\epsilon_{\mathbf{k}}= \frac{\hbar^2 \mathbf{k}^2}{2m}$. In the interaction term, $\mathbf{K}= \{ \mathbf{k}, \mathbf{k}',\mathbf{q} \}$ and
$U_{\mathbf{kk'}}$ is the Fourier transform of the finite short-range potential that models the interparticle interaction $U_{ \mathbf{k} \mathbf{k}'} = \frac{1}{V}\int  e^{i (\mathbf{k} - \mathbf{k}') \cdot \mathbf{r} }U(r) d^3r$. The finite-range potentials here studied are a square-well, typically used as a first approach to model the interactions \cite{Ketterle-review}, a purely decaying exponential, a classic nuclear model \cite{Rarita}, the Yukawa potential \cite{Rowlinson-yukawa}, and a Van der Waals potential, a typical atomic interaction tail \cite{strinatibcs}. These four models, representative of short-range interatomic potentials, are expressed in terms of a characteristic energy scale $V_0$ and its spatial finite range $R$. 
In our calculations we keep the BCS interaction between atoms of different species only. It can be shown that the Hartree and Fock terms, in the ground state, do not modify the essential physics of the crossover transition, not only in the contact interaction but for the finite-range interaction case as well \cite{Eleazar}.
 After substituting the BCS ansatz $|\Psi_{\mathrm{BCS}}\rangle= \Pi_{\mathbf{k}} \left(u_{ \mathbf{k}}+v_{ \mathbf{k}} a_{ {\mathbf{k}} \uparrow}^{\dagger} a_{- {\mathbf{k}} \downarrow }^{\dagger} \right) |0 \rangle $, and performing the standard variational procedure in the grand potential function $\Omega= \langle \Psi_{\mathrm{BCS}}| H-\mu N| \Psi_{\mathrm{BCS}} \rangle$,
one obtains $2u_{\mathbf{k}}^2=1+ (\epsilon_{\mathbf{k}}-\mu)/ E_{\mathbf{k}}$ and $2v_{\mathbf{k}}^2=1- (\epsilon_{\mathbf{k}}-\mu)/ E_{\mathbf{k}}$, where $E_{\mathbf{k}}=\sqrt{ (\epsilon_{\mathbf{k}}-\mu)^2+ \Delta_{\mathbf{k}}^2 }$. The energy gap $\Delta_{\mathbf{k}}$ and chemical potential $\mu$ satisfy the coupled equations,
\begin{equation}
\begin{aligned}\Delta_{\mathbf{k}} & =-\frac{1}{V}\sum_{\mathbf{k}'}U_{\mathbf{k}\mathbf{k}'}\frac{\Delta_{\mathbf{k}'}}{2E_{\mathbf{k}'}}\,,\\
N & =\sum_{\mathbf{k}}\left(1-\frac{\epsilon_{\mathbf{k}} - \mu}{E_{\mathbf{k}} }\right),
\end{aligned}
\label{coupledeqs}
\end{equation}
with $N$ the total number of particles. 
Because of the $s$-wave symmetry, the dependence of $\Delta_{\mathbf{k}}$ on ${\mathbf{k}}$ is on its magnitude $k$ only and, certainly, on the parameters $V_0$ and $R$. It is very important to emphasize here that for finite-range potentials the equation for $\Delta_{\mathbf{k}}$ does not require the standard renormalization procedure used in the case of the contact interaction approximation, to warrant convergence \cite{strinatibcs}. In fact, such a divergence is an artifact associated with the contact approximation itself \cite{SCaballero}. Once Eqs. (\ref{coupledeqs}) are solved, one can calculate the functions $u_{\mathbf{k}}$ and $v_{\mathbf{k}}$, on which the correlation functions are expressed.\\

As broadly established in the literature, the BEC-BCS crossover for the contact interaction is generally analyzed in terms of the $s$-wave scattering length $a$, which, besides being the effective potential amplitude, signals the emergence of a bound state as it diverges \cite{Feshbach-review,strinatibcs,Ketterle-review}. In contrast to the contact interaction, the expressions for the finite-range potentials here considered do not depend on $a$. This, however, is not an obstacle to investigate the mean-field many-body physics in terms of such a length. The key aspect is to have a relationship of $a$ in terms of the potential parameters $V_0$ and $R$ in order to reveal the presence of bound states or dissociated pairs. In some cases, we may have an analytic expression of $a$ as a function of $V_0$ and $R$, but in general, we can find such a relationship by numerically solving the two-body Schr\"odinger equation in the low-energy limit \cite{Jeszenszki}. We make this correspondence for the proposed interaction potentials in order to investigate the correlations along the BEC-BCS crossover, in terms of $a$. Based on the results of Refs. \cite{Eleazar,Parish_finite-range}, in this Letter we consider a representative value for the range $k_F R = 0.1$, with $k_F = (3 \pi^2 N/V)^{1/3}$ the Fermi momentum. This choice is consistent with the hypothesis of a weakly interacting gas; we shall comment below about larger values of $R$. \\

 To investigate the spatial large-distance behavior of the correlation functions for the finite-range potentials here considered, see Fig. \ref{fig_large-distance_correlation_length}, we recall that the density-density correlation functions are defined as \cite{strinatibcs,JCO1}, 
\begin{equation}
G_{\sigma \sigma '} (\mathbf{x}, \mathbf{r}) = \langle \hat n_\sigma(\mathbf{x}) \hat n_{\sigma '}(\mathbf{r}) \rangle -\langle \hat n_\sigma(\mathbf{x}) \rangle \langle  \hat n_{\sigma '}(\mathbf{r}) \rangle.
\end{equation}
where as stated above, $\sigma$ and $\sigma '$ identify different hyperfine spin states. The density operator for the spin component $\sigma$ is given by $\hat{n}_{\sigma}(\mathbf{x}) = \hat{\psi}_{\sigma}^{\dagger}(\mathbf{x}) \hat{\psi}_{\sigma}(\mathbf{x})$, 
where $\hat{\psi}_{\sigma}(\mathbf{x})$ are the usual field operators of a uniform gas. 
For a balanced mixture, the condition of equal population $N_{\uparrow} = N_{\downarrow}$ implies
$G_{\uparrow \downarrow} (\mathbf{r}) = G_{\downarrow \uparrow } (\mathbf{r})$ and $G_{\uparrow \uparrow} (\mathbf{r}) = G_{\downarrow \downarrow } (\mathbf{r})$.\\

It can be straightforwardly shown that the correlation function for antiparallel spins is \cite{JCO1,strinatibcs,Ketterle-review}, 
\begin{equation}\label{eq_Guu}
G_{\uparrow \downarrow}(\mathbf{r}) = |g_{\uparrow \downarrow}(\mathbf{r})|^2,
\end{equation}
where,
\begin{equation}\label{eq_fouriert_ud}
g_{\uparrow \downarrow} (\mathbf{r}) = \frac{1}{(2 \pi)^3} \int d^3 k \; e^{i \mathbf{k} \cdot \mathbf{r}} \; u_k v_k .
\end{equation}
And, correspondingly, the correlation function for parallel spins is given by,
\begin{equation}
G_{\uparrow \uparrow} (\mathbf{r}) = \frac{n}{2} \delta^{(3)}(\mathbf{r}) - |  g_{\uparrow \uparrow}(\mathbf{r}) |^2,
\end{equation}
where
\begin{equation}\label{eq_fouriert_uu}
g_{\uparrow \uparrow} (\mathbf{r}) = \frac{1}{(2 \pi)^3} \int d^3 k \; e^{i \mathbf{k} \cdot \mathbf{r}} \;  v_k^2 .
\end{equation}
An associated quantity in the BCS theory is the Cooper pairs wavefunction, 
\begin{equation}\label{eq_ecuacion_del_par}
\phi_{\text{BCS}}(r) = \frac{1}{(2 \pi)^3} \int d^3 k \; e^{i \mathbf{k} \cdot \mathbf{r}} \; \frac{v_k}{u_k}.
\end{equation}
As we will illustrate below, the Cooper-pair density probability distribution $|\phi_{\text{BCS}}(r)|^2$ shares the same asymptotic features of
the correlation functions.\\

\begin{figure}[h!]
\centering
\includegraphics[width=\linewidth]{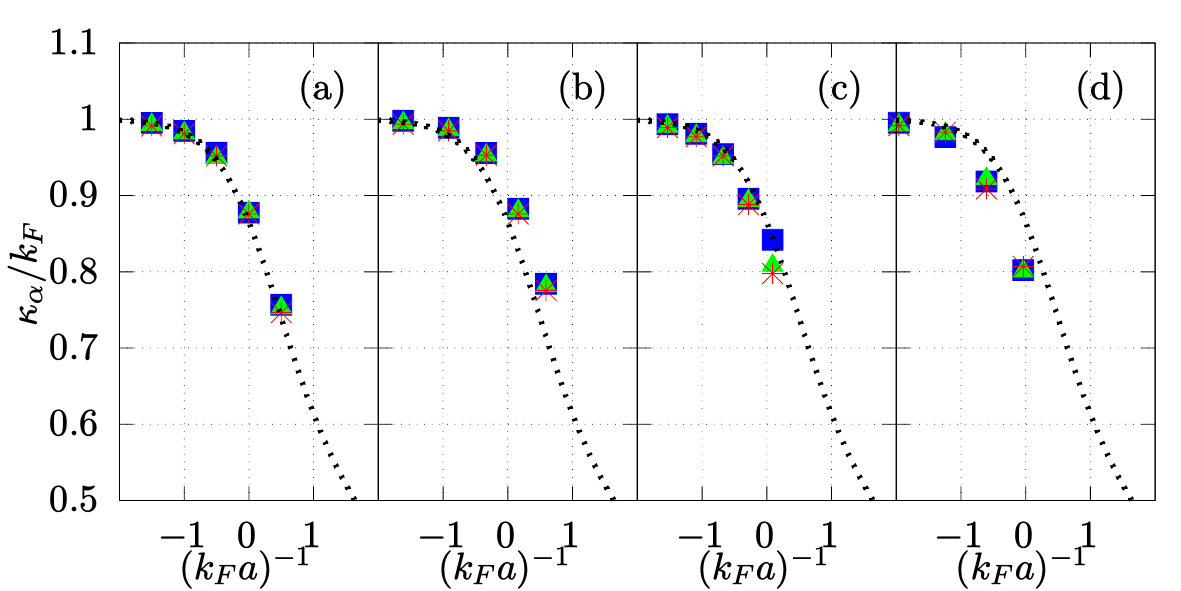}
\caption{ Wave vectors $\kappa_\alpha$ for the (a) square well, (b) exponential, (c) Yukawa, and (d) Van der Waals potentials. Symbols are the same as in Fig. \ref{fig_large-distance_correlation_length}. The dotted (black) line corresponds to the contact potential case reported in \cite{JCO1}.}
\label{fig-wavevector}
\end{figure}

As stated above, we first numerically solve the gap and number equations, Eqs. (\ref{coupledeqs}), to obtain $v_{\bf k}$ and $u_{\bf k}$. Then, by numerical Fourier transforms, we calculate the correlation functions $G_{\uparrow \uparrow}(\mathbf{r})$, $G_{\uparrow \downarrow}(\mathbf{r})$ and the density pair probability $|\phi_{\text{BCS}}(r)|^2$, see Eqs. (\ref{eq_Guu}) - (\ref{eq_ecuacion_del_par}). This is a difficult task, specially for finding their asymptotic behavior for long distances $ k_F r \gg 1$. While in the deep BCS and BEC regimes, $(k_Fa)^{-1} \to \pm \infty$, the numerical convergence does not allow to make definite conclusions, fortunately the best fittings can be found for values near unitarity, namely $- 3 \lesssim (k_F a)^{-1} \lesssim 3$, which covers experimentally accessible values \cite{schunckdetermination,Ketterle-review,strinatibcs,Giorgini-Stringari}. Figures \ref{fig_large-distance_correlation_length} and \ref{fig-wavevector} are the main result of this Letter, that we find them very impressive: the correlation functions and the pair probability distribution, for large distances $k_Fr \gg 1$, can be cast as  distributions of the form
\begin{equation}\label{correlations}
\rho_\alpha(r) \approx \frac{\rm const}{r^2} e^{-r/\xi_\alpha} {\cal P}(\kappa_\alpha r + \varphi_\alpha)\>,
\end{equation}
where $\alpha = \uparrow\downarrow,\uparrow\uparrow, \text{BCS}$ and 
${\cal P}(\kappa_\alpha r + \varphi_\alpha)$ a periodic function to be determined and which, in principle, depends on the particular model potential. In Fig. \ref{fig_large-distance_correlation_length} we show fittings of $\xi_\alpha$ for $G_{\uparrow \downarrow}(\mathbf{r})$ in (blue) squares, $G_{\uparrow \uparrow}(\mathbf{r})$ in (green) triangles and $|\phi_{\text{BCS}}(r)|^2$ in (red) asterisks, for the three finite-range potentials considered. As stated above, the outstanding finding is the identification of an universal expression for the correlation length $\xi_\alpha = \xi = (\hbar^2/4 m \epsilon_{\text{spec}})^{1/2}$, with $\epsilon_{\text{spec}}$ given by Eq. (\ref{espec}), for both correlation functions and the pair-particle distribution, which is shown as a solid (purple) line in Fig. \ref{fig_large-distance_correlation_length}. Note that in such a figure we also include the contact interaction case, $U(r) = 4 \pi \hbar^2 a \delta({\bf r})/m $, in a dotted (black) line. \\

In Fig. \ref{fig-wavevector} we plot the fitting of the long distance wave vector oscillations $\kappa_\alpha$. Again, for each interatomic potential, the three functions $G_{\uparrow \uparrow}(\mathbf{r})$, $G_{\uparrow \downarrow}(\mathbf{r})$ and   $|\phi_{\text{BCS}}(r)|^2$ oscillate with the same wavelength, along the crossover, as a function of $(k_F a)^{-1}$. For this quantity, although we do not have a function that fits it, we find the general conclusion that in the BCS side the wave vector tends to the Fermi momentum, as expected, $\kappa \to k_F$, then it falls down to zero towards the BEC side. As can be seen in the figure, we were not able to fit the oscillations in the BEC side since they are severely arrested by the rapidly falling correlation length. As a reference, the contact interaction case do explicitly show that the wavelength of the oscillations grows with no bound as $(k_F a)^{-1} \to + \infty$. This oscillatory behavior can be identified with the many-body configuration. In the BCS limit the wave vector is similar to the Fermi wave number, in agreement with the weakly interacting regime. In the BEC limit the wave vector decreases allowing the pair wave function and the opposite spins correlation function to converge to a bound state distribution. Although we do not present a figure to demonstrate it, we find that the correlation functions $G_{\uparrow \uparrow}(\mathbf{r})$ and $G_{\uparrow \downarrow}(\mathbf{r})$ oscillate out of phase, just as in the contact interaction case \cite{JCO1}, indicating the already identified nested structure of the different spin species, as statistically ``observed'' from any atom. \\

Let us now turn to the insight of identifying $\epsilon_{\text{spec}}$ as the dominant energy scale at large distances. First, this behavior has already been observed for the contact potential in 3D. In Ref. \cite{JCO1} this was done by extending the Fourier transform integrals of Eqs. (\ref{eq_fouriert_ud}), (\ref{eq_fouriert_uu}) and (\ref{eq_ecuacion_del_par}) to the complex plane, allowing for a direct identification of an exponential decay in terms of $\xi$. Additionally, and very recently, in Ref. \cite{Strinati-2022} it was shown that for the one-body density matrix of the mean-field contact-interaction Fermi gas, indeed, $\xi$ is the characteristic length scale. On the other hand, for contact interactions in 2D, it has been found that $\xi$ is the exact characteristic length of the correlation functions across the whole crossover \cite{JCO2}. Moreover, also in Ref. \cite{JCO1}, the universal oscillatory wave vector $\kappa$, Fig. \ref{fig-wavevector}, was found for the contact interaction potential. \\

\begin{figure}[h!]
\centering
\includegraphics[width=\linewidth]{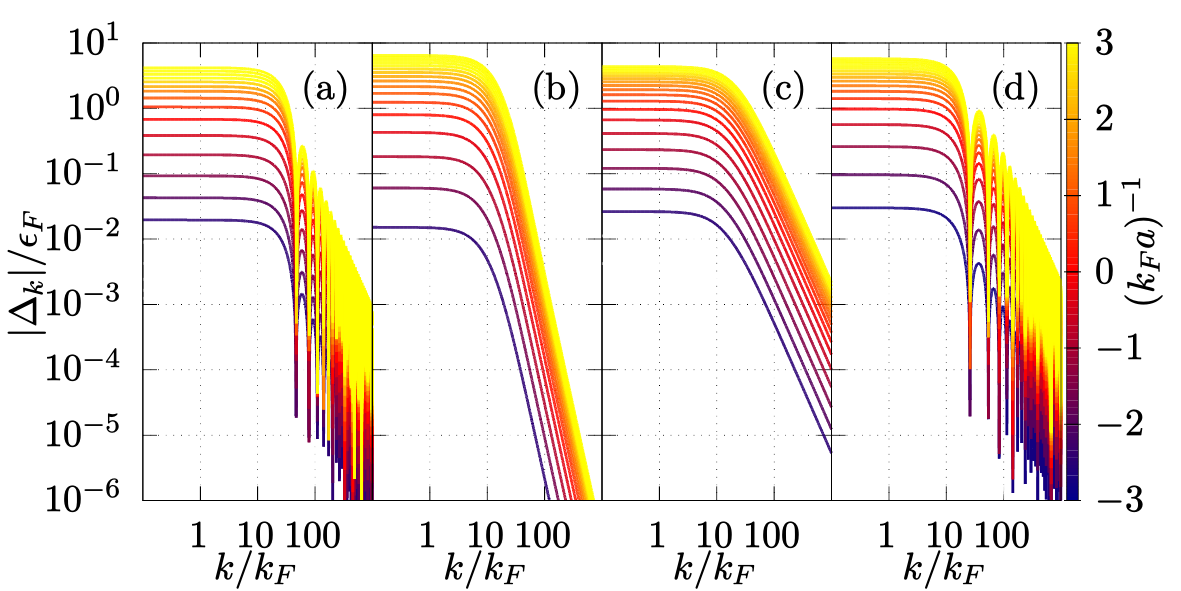}
\caption{(Color online) Dimensionless gaps $|\Delta_{k}|/\epsilon_F$, scaled with the Fermi energy $\epsilon_F = \hbar^2 k_F^2/2m$, for the (a) square well, (b) exponential, (c) Yukawa, and (d) Van der Waals potentials. Each line corresponds to different values of the $s-$wave scattering length $a$. Both axes have a logarithmic scale.}
\label{Figgap}
\end{figure}

To envision the meaning of the so-called spectroscopic energy $\epsilon_{\text{spec}}$ we follow the observation of Ketterle et al. \cite{Ketterle-review,schunckdetermination} who established that this is a threshold dissociation energy to break a pair, certainly susceptible of being measured. Those authors point out to the fact that, besides the balanced two-component mixture with $\uparrow$ and $\downarrow$ spins, a third hyperfine spin state can be used to capture the energy arising from a pair in the BCS state when it dissociates. It is important to stress that this third state has experimental relevance since typically a radio frequency signal with a definite magnitude, arising from the transition to such an additional state, confirms the identity of the pairs along the crossover. The energy difference for an excitation minus the energy associated with the third hyperfine state $\epsilon_{3}$ may be generally written as \cite{schunckdetermination},
\begin{equation}
\Delta E-\epsilon_3= \sqrt{ (\epsilon_{\mathbf{k}} - \mu)^2 + |\Delta_{\mathbf{k}}|^2}+ (\epsilon_{\mathbf{k}} -\mu) \>.
\end{equation}
Then, one can readily find that the minimum value for the pair dissociation is found at $\mathbf{k} = 0$, see Fig. \ref{Figgap}, thus yielding $\epsilon_{\text{spec}}$ as expressed in Eq. (\ref{espec}). It is evident that this derivation is independent of the interaction potential used and, for sure, whether is of finite range or of contact nature. However, the values of the gap $\Delta_0$ and the chemical potential $\mu$ do certainly depend on the details of the interaction potential. \\

The universality of the correlation length partly rests on the fact that the minimum value of the dissociation energy occurs at $k = 0$ where, in turn, the gap $\Delta_{\bf k}$ has its maximum value. This is seen in Fig. \ref{Figgap} where the gap is plotted as a function of $k$ and $a$ for the analyzed finite-range potentials. One can see from this figure that, either on the BCS side ($a<0$), or on the molecular BEC sector ($a>0$), the gap takes its maximum value at $k=0$, and then it decays as $k$ grows. We recall here that in the contact interaction case the gap is constant. A concise observation regarding the square well and Van der Waals potentials is that oscillations for large values of $k$ arise as a result of the discontinuity at $R$, that is, for small values of $r$.  Although not shown here, as the range $R$ of the potential is diminished the region where the gap remains almost constant increases. On the other hand, as $R$ is increased, while $\xi$ as given by $\epsilon_{spec}$, Eq. (\ref{espec}), remains a universal result, its value starts to depart from the value of the contact interaction, shown in Fig. \ref{fig_large-distance_correlation_length} with the dotted (black) line. These results evidently indicate that, as the finite range $R$ of the potential becomes smaller, the asymptotic large-distance behavior of the contact interaction becomes a better approximation for a true finite short-range interatomic interaction potential.\\

To conclude we point out that, while the meaning of the spectroscopic energy $\epsilon_{spec}$
and its associated length is clear, the fact that this is the correlation length of all the
two-body functions is not at all evident. Furthermore, although in hindsight one may argue that the universal dependence of the correlation on
 $k_Fa$ is a consequence of the mean-field BCS theory, having a simple structure at small wavevectors where the details of the
interaction potential become irrelevant, there still remains to find a deeper and general argument to explain
such a result. The findings of this article may motivate experiments for measuring correlation functions and verify whether their asymptotic behavior is indeed similar for different atomic systems or not.

\acknowledgments{This work was partially funded by grant  IN108620 DGAPA (UNAM). GADC and JCOJ acknowledge CONACYT scholarship.}

\bibliography{bibliography-FR}

\end{document}